# New investigations of the porous dielectric detectors[*]

## M.P. Lorikyan

Yerevan Physics Institute,
Brothers Alikhanian str.2,
Yerevan, 375036 Armenia
Tel/Fax: (3741) 342838
E-mail: lorikyan@moon.yerphi.am
  lorikyan@star.yerphi.am

## Abstract

The results of the study of the multiwire porous dielectric detector and microstrip porous dielectric detector filled with porous CsI are presented. Detectors were exposed to $\alpha$-particles with energy of 5.46 MeV and X-rays with energy of 5.9 keV and operated with a constant applied voltage. It is found that right after thermal deposition of the porous CsI layer the detector's performance is unstable, it has poor spatial resolution. However in a course of time the performance becomes stable and it acquires high spatial resolution.

---

[*] The work is supported by the International Science and Technology Center

# Introduction

The functioning of the porous dielectric detector is based on the phenomenon of the drift and multiplication of electrons in the porous media under the action of the external electric field. In a case when the drift and multiplication of electrons (EDM) occur under action of internal electric fields, the emission of the electrons into vacuum (Anomalous Secondary Electron Emission - Malter effect) [1, 2] is inertial and non controllable and the secondary electron emission coefficient for relativistic single particles $\delta_{anom} \approx 1$ [3, 4]. This phenomenon has not found application for radiation registration. In 1970-s intensive researches of drift and multiplication of electrons in porous dielectrics and their emission in vacuum under influence of an external electric field were carried out in the Yerevan Physical Institute. The idea was in the following: Malter effect arises and it proceeds under the action of fields of the large surface and space charges, therefore EDM process is non controllable and inertial. In an external electric field, if the emitter is irradiated by a not intensive beam of particles, these charges do not arise, therefore was anticipated, that processes of drift and multiplication of electrons will be non inertial and governable by the external electric field, and that $\delta$ for relativistic particles will considerably exceed $\delta_{anom}$. The use of external field made it possible to investigate the drift and multiplication of electrons in porous dielectrics (EDM in PDs) for thick layers of porous dielectrics [5-10]. Researches have shown, that EDM in PDs and the electron emission in vacuum under influence of an external electric field are non inertial and controllable, and that the secondary electron emission coefficient for minimally – ionizing single particles reaches $\delta_{cont}=250$ [6]. Then it was shown, that in case of α-particles with energy of $\approx 5.46$ MeV, $\delta_{con}$ reaches several thousands [9]. Later the secondary electron emission in porous dielectrics placed in the external electric field was investigated by C. Chianelli, et al. [11] and R. Cheehab et al. [12], who confirmed that this phenomenon indeed is non inertial and controllable and that the secondary electron emission factor for high energy particles is large.

In 1978 the same Yerevan group developed and investigated porous multiwire detectors (MWPDD) [13-18]. In these investigations it was found, that when the MWPDD was working in the DC mode, it was unstable and the efficiency of particle registration dropped to very low values in less than one hour. To stabilize the performance of porous dielectric detectors, periodically after a short operating time (several milliseconds) the applied voltage was turned off, then a working pulse of inverse polarity was applied on the detector, i.e. these detectors maintained a stable performance in a pulsed mode of operation [19-21]. In the pulsed operation mode the coordinate resolution of these detectors is 125 μm [18, 22]. The time resolution of MWPDD is 60 ps [23] b and detection efficiency of minimally-ionizing particles is 100% [24]. More full information on problem of porous detectors is given in reference [25].

The preliminary results of investigation of a porous microstrip detector are presented in [25]. Later Lorikyan continued these investigations [26, 27] and has shown, that porous detectors operation in the continuous DC mode is also stable.

Qualitatively, the mechanism of the EDM in porous dielectric media under the action of an external electric field can be represented in the following manner: the primary particle knocks out electrons from valence band of material pore walls into a conduction band. The holes remain in the valence band. In the conduction band electrons with energies higher than the electron affinity $\chi$ of the pore surfaces escape from the pore walls and are accelerated in the pores by the external electric field. The accelerated electrons induce in the low depths of pore walls the same processes as initial particles. These processes repeat in the second, third, etc. generations of electrons. When in each act of electron collision with pore walls the secondary electron emission factor $\delta > 1$, an avalanche multiplication of electrons (holes) in the porous medium may take place. The process of EDM and avalanche of electrons are schematically shown in Fig. 1. Electrons and holes move in opposite directions and during the increase of the distance between them a charge is induced on electrodes. This charge is proportional to a potentials difference between points, which they have reached. The value of $\chi$ in the dielectrics is sufficiently high and thermal electrons cannot take part in the emission process, therefore the initiation of cascade processes of the electron multiplication is practically impossible. So, the high electron multiplication factor observed in porous dielectrics is possibly explained by the decrease of $\chi$, as in the case of the P-type semiconductors, where the electric field decreases the $\chi$ to negative values if the surface is covered by monoatomic layer of Cs or $Cs_2O$ [28]. In our case the Cs atoms are created by the dissociation of CsI molecules. When $\chi$ decreases, the thickness of the layer (escape depth) from which electrons reach the surface and escape into vacuum increases [28]. All these effects provide a high factor of secondary electron emission from pore walls. However numerous defects in the crystalline structure exist on the surfaces and in the walls of porous dielectrics. These defects are charge carrier capturing centers and lead to a formation of a large space and surface charges. The direction of the electric field of the latter is opposite to that of the external field, that is why the speed of the electron drift and the electrons multiplication factor decrease. Because of that, the loss of the charge carriers increases and also their scattering on the walls is more intensive, thus the time stability, the spatial and time resolutions become poor. So, detectors with dielectric working media having a notable density of charge carrier capturing centers will have a poor spatial, energy and time resolutions and unstable performance. Beside this kind of traps, there exist also impurity traps, that is why the working medium purity is very crucial for normal operation of porous detectors [29].

In recent works [24, 25] the multiwire porous dielectric detectors (PDD) were prepared by paying more attention to the cleanness of the technology. Detectors were investigated in the DC mode. It was found that, some hours after manufacturing of porous CsI layer the performance of porous detectors was unstable and it had a low spatial sensitivity. Thus inside the porous CsI and on the surfaces of its walls the densities of the surface and spatial charges were possibly rather large. But in due course of time detectors acquire stability and a high coordinate resolution and high particles detection efficiency. After that PDDs maintained stable performance and had high counting characteristics, which did not change even after being turned off for long time. This means, that the densities of the surface and spatial charges in porous CsI have dropped drastically and also the number of charge carrier capturing centers decreased significantly and the influence of polarization effects on the drift and multiplication of the electrons became insignificant.

In this paper the results of the study of the MWPDD and MSPDD, filled with porous CsI, are presented. Detectors are exposed to α-particles of energy of 5.46 MeV and X-rays of energy of 5.9 keV. These detectors were placed after assembly in ambient Ar inside the vacuum chamber which was then pumped out to $\approx 7 \times 10^3$ Torr. Investigations of detectors started one hour after their putting in the vacuum chamber. All these procedures and measurements are made at room temperature $\approx 18^0$C. The time stability, spatial resolution and radiation registration efficiency were measured in DC mode.

For quick testing of the spatial sensitivity of the detectors and determining the upper limit of the spatial resolution, which is restricted by the transverse sizes of electron shower Z the the following method was used: without collimation of the sources and with a constant registration threshold $V_{thr}$, in the same experimental run both the number of particles (N(I)) detected only by one anode wire (strip) (or corresponding efficiency of registration of particles η(I)) and the number of particles (N(C)), which are detected simultaneously by the two adjacent anode wires (strips) (or probability of registration of such events η(C)), were measured. It is evident that the ratio N(I)/N(C) = Ss (spatial sensitivity of porous layer) defines the spatial resolution of detector. When the width of the electron shower is much smaller than the distance between the adjacent anode wires, each particle is detected in general by only one anode wire (strip) and Ss >> 1. In the opposite case the particles are detected in general simultaneously by two adjacent anode wires (strips) and Ss ≤ 1. When the η(I)=1 and Ss >>1, the coordinate resolution of detector $\sigma_x = b/(2\sqrt{3})$.

The purity of CsI used in manufacturing the detectors was 99.99 %. The fast current amplifiers with a rise time of 1.5 ns, 250-Ω input impedance, and the conversion ratio of 30 mV/μA were used for the amplification of detector's pulses. In the intervals between measurements of each point, the detectors were exposed to radiation and a voltage was applied to them, however in experiment lasting for many days the detector was switched off on weekends and also for 16 hours at nights. In all cases, measurement errors (not indicated in figures) are only statistical. The intrinsic noises of detectors and amplifier were $\approx 0.1$-$0.2$ s$^{-1}$.

## 1. Description of porous detectors

The schematic view of MWPDD is shown in Fig. 2. The 25 μm diameter anode wires are made of gilded tungsten and are spaced at b = 0.25 mm. The cathode is made of 60 μm Al foil. The porous CsI layer is prepared by thermal deposition in an Ar atmosphere [30] at a pressure of p=3 Torr. Initially a CsI layer was sputtered onto the cathode, afterwards, the frame with the anode wires is mounted on the cathode. The detectors had gap of 0.5 mm.

In the case of detection of α-particles a porous CsI layer just after deposition had a thickness of 0.82 mm and a density of $\rho/\rho_0$=0.35%, where $\rho_0$ is a CsI monocrystal density. The sensitive area of this detector was 22.5 ×22.5 mm$^2$. α-particles with intensity of 476 min$^{-1}$cm$^{-2}$ pass to the porous medium through the gaps between the anode wires.

In the case of detection of X-rays a porous CsI layer of MWPDD just after deposition had a thickness of 0.75 ±0.05 mm and a density of $\rho/\rho_0$ = 0.67 %. The sensitive area of this MWPDD was of 25 ×25 mm$^2$. X-rays of intensity of 630 min$^{-1}$cm$^{-2}$ were entering onto the porous CsI through the Al foil cathode.

The schematic view of MSPDD is given in Fig. 3. Golden strips were lithographically deposited onto the glass-ceramic plate. The width of strips was 20 μm. The distance between their centers was 100 μm, and the gap between the strips and the cathode electrode was 0.5 mm. The cathode was made of 0.7 transparency micromesh. The sensitive area was of 22 ×22 mm$^2$. Immediately after the deposition the thickness and the density $\rho/\rho_0$ of porous CsI layer were 0.84 ±0.05 mm and 0.37 % respectively. Initially CsI layer was sputtered onto the plate with microstrips. Afterwards, the cathode was mounted in such a manner that the porous layer compacted from the thickness of 0.84 mm to 0.5 mm, i.e. the porous CsI layer was compacted 1.7 times. α-particles of intensity of 630 min$^{-1}$cm$^{-2}$ were falling from the side of the micromesh.

## 2. Experimental results.

### 2.1. Investigation of the Multiwire porous dielectric detector (MWPDD).

Figure 4 shows the dependences of the number of detected α-particles, $N_\alpha(I)$ (squares) and $N_\alpha(C)$ (triangles), on the voltage U. The intensity of α-particle was 676 min$^{-1}$cm$^{-2}$. Measurements were performed in one hour after assembling the MWPDD. One can see, that $N_\alpha(I)$ in the beginning grows up to $N_\alpha(I) \approx 300$ (the efficiency of registration $\eta_\alpha(I)$=0.1), but then decreases. $N_\alpha(C)$ also grows, but reaches a plateau of $N_\alpha(C) = 2400$ (probability of registration of each particles by the two adjacent anode wires $\eta_\alpha(C)$=0.86). For small U the spatial sensitivity Ss is close to unity but with increasing U it decreases and in the region of the plateau of $N_\alpha(C)$ Ss $\ll 1$. So, the spatial resolution of the detector is worse than the anode-wire spacing 250 μm. The MWPDD's time stability, determined immediately after these measurements were completed, is given in Fig. 5. From it is clear that $N_\alpha(I)$ and $N_\alpha(C)$ in 30 minutes drop from $N_\alpha(C)=2200$ practically to zero, i.e. the detector performance is unstable. Thus the density of defects in the crystalline structure of the porous layer is very high.

After completion of these measurements the MWPDD was switched off and measurements were resumed in 19 hours. The U-dependences of $N_\alpha(I)$ (squares) and $N_\alpha(C)$ (triangles) are shown in Fig. 6.

Comparing Fig. 4 and Fig. 6 we see that first the working voltage U is the latter case has shifted by ~150 V to the right, second $N_\alpha(I)$ in difference from the previous case increases rapidly and reaches a plateau for $N_\alpha(I)=2900$, while $N_\alpha(C)$ for a low U is very small and increases insignificantly for U-s below the one, corresponding to the plateau of $N_\alpha(I)$, and in the plateau region is only 6% of $N_\alpha(I)$, i.e., in the region of the working voltage Ss $\gg 1$. Thus in course of time the MWPDD's working voltage increases, the detector acquires high spatial sensitivity and a spatial resolution becomes σ=72 μm since $\eta_\alpha(I)=1$.

Results of measurements of the time-stability of $N_\alpha(I)$ and $N_\alpha(C)$ observed immediately after obtaining previous results are shown in Fig. 7. It is easy to see the detector performance is stable.

The time stability of the detection efficiencies of α-particles $\eta_\alpha(I)$ (squares) and probability of the simultaneous registration of the same α-particles by two adjacent anode wires $\eta_\alpha(C)=N_\alpha(C)/N_{0\alpha}$ (triangles) of this MWPDD for 130 days are given in Fig. 8. One can clearly see that $\eta_\alpha(I)$ during all that time with high accuracy had a constant value, while the $\eta_\alpha(C)$ drops steeply on the 124-th day. Note that on the second day of measurements of the time stability U=963 V, and on 130-th day had grown to 1100 V. The MWPDD was switched off every day for 16 hours, from 47-th to 56-th day and always in weekends. Also we have observed, that a switched off detector, which remained in the vacuum chamber for 3 months, had the same characteristics as before that. Thus, the counting characteristics of MWPDD's did not change also if they were switched off for a long time.

The results of investigation of the time-stability of MWPDD exposed to X-rays in the period from the 3-rd day to the 17-th day are shown in Fig. 9. The detection efficiency of the X-quantum by only one anode wire $\eta_x(I)$ and probability of its registration simultaneously by two adjacent wires $\eta_x(C)=N_x(C)/N_{0x}$ are marked by squares and triangles, respectively. One can see, that within the experimental errors $\eta_x(I)$ remains constant during all that time while $\eta_x(C)$ decreases for about three times. Thus the performance of MWPDD in the case of X-rays, when the energy loss is 1000 times less than in case of α-particles, is also stable and it has high spatial resolution $\sigma_x$.